**Influenza-associated mortality for different causes of death during the 2010-2011 through the 2014-2015 influenza seasons in Russia**

Edward Goldstein[1,*]

1. Center for Communicable Disease Dynamics, Department of Epidemiology, Harvard TH Chan School of Public Health, Boston, MA 02115 USA

*.  Email: egoldste@hsph.harvard.edu

**Abstract**

Background: There is limited information on the volume of influenza-associated mortality in Russia.

Methods: Using previously developed methodology (Goldstein et al., Epidemiology 2012), we regressed the monthly rates of mortality for respiratory causes, circulatory causes, and for certain infectious and parasitic diseases (available from the Russian Federal State Statistics Service (Rosstat)) linearly against the monthly proxies for the incidence of influenza A/H3N2, A/H1N1 and B (obtained using data from the Smorodintsev Research Institute of Influenza (RII) on levels of ILI/ARI consultations and the percent of respiratory specimens testing positive for influenza A/H3N2, A/H1N1 and B), adjusting for the baseline rates of mortality not associated with influenza circulation and temporal trends.

Results: For the 2010/11 through the 2014/15 seasons, influenza circulation was associated with an average annual 11106 (95% CI (4321,17892)) deaths for circulatory causes, 4552 (3744,5360) deaths for respiratory causes, and 343 (63,624) deaths for certain infectious and parasitic diseases, with influenza A/H3N2 being responsible for the majority of influenza-associated respiratory deaths and influenza B making a substantial contribution to influenza-associated mortality for circulatory causes.

Conclusions: Influenza circulation is associated with a substantial mortality burden in Russia, particularly for circulatory deaths. Those results support the potential utility of influenza vaccination (with the role played by influenza B in circulatory mortality pointing to the benefit of quadrivalent influenza vaccines), as well as of administration of antiviral drugs for high-risk individuals during periods of active influenza circulation.

**Introduction**

Annual epidemics associated with the circulation of influenza A/H3N2, A/H1N1 and B take place in Russia [1]. Such epidemics are known to result in a substantial toll of mortality in other countries in the Northern Hemisphere, e.g. [2-7], including mortality for circulatory causes [2,3], with influenza additionally known to be associated with a variety of cardiovascular manifestations [8]. At the same time, there is limited information on the national burden of influenza-associated mortality in Russia. Estimates of the rates of influenza-associated mortality in Russia for select influenza seasons exist on a regional level, particularly for the city of St. Petersburg [9]. At the same time, the rates of influenza-associated mortality vary both spatially, as well as year-to-year. For example, while influenza A, particularly influenza A/H3N2 is known to be a major source of excess mortality [10], the 2017/18 season in Europe saw high levels of excess mortality associated with the circulation of influenza B/Yamagata strains [11], with those strains not being included in trivalent influenza vaccines. Additionally, the overall rates of death for circulatory diseases in Russia are high [12], and there is evidence that influenza B plays an important role in influenza-associated circulatory mortality, e.g. [13], as well as Table 3 in [2] (comparison of regression coefficients for circulatory vs. respiratory mortality). It is important therefor to consider national data from a number of influenza seasons to address the uncertainty regarding the national burden of mortality associated with the circulation of the major influenza (sub)types (A/H3N2, A/H1N1 and B) in Russia for different causes of death, including respiratory and circulatory diseases. Such estimates are important for planning mitigation efforts, including vaccination and administration of antiviral medications to certain population groups during active circulation of the different influenza (sub)types.

In our earlier work [2,3,14], we introduced a new method for estimating the burden of severe outcomes associated with influenza and the respiratory syncytial virus (RSV), designed to address several limitations of some of the previously employed inference models. An important feature of that approach is the use of RSV and influenza A/H3N2, A/H1N1, and B incidence proxies that are expected to be linearly related to the population incidence of those viruses; such proxies for influenza (sub)types combine data on medical consultations for influenza-like illness (ILI) with data on the testing of respiratory specimens from symptomatic individuals [2,3,7]. We've applied those incidence proxies to estimate the rates of influenza-associated mortality stratified by age/cause of death in the US [2,3], with the corresponding method being later adopted for the estimation of influenza-associated mortality in other countries as well, e.g. [5-7]. In this paper, we derive the analogous proxies for the weekly/monthly incidence of influenza A/H3N2, A/H1N1, and B in Russia based on the surveillance data from the Smorodintsev Research Institute of Influenza (RII) [15]. We then apply the inference model from [2,3,14] to relate those incidence proxies to the monthly data from the Russian Federal State Statistics Service (Rosstat) [16] on the rates of death for respiratory causes, circulatory causes, and deaths for certain infectious and parasitic diseases to estimate the corresponding mortality rates associated with the circulation of the major influenza (sub)types during the 2010/11 through the 2014/15 influenza seasons in Russia.

## Methods

### *Data*
Monthly data on mortality for certain major causes of death in Russia are obtained from [16]. We'll concentrate on respiratory deaths, circulatory deaths, and deaths for certain infectious and parasitic diseases (with about 60% of the latter deaths in 2018 being HIV-related, and another 25% being TB-related, [12]). Influenza is known to be associated with mortality for other causes as well, including neoplasms and metabolic diseases [2,9]; however finer (e.g. weekly) mortality data are needed to ascertain the signals from influenza circulation in the corresponding time series.

Weekly data on the rates of ILI/ARI consultation per 10,000 individuals in Russia are available from [15] (with some of the data during the study period missing in [15] and can be found in the Russian language version of the Smorodintsev Institute Surveillance database [17]). Refs. [15,17] also contain the weekly data on the percent of respiratory specimens from symptomatic individuals that were RT-PCR positive for influenza A/H1N1, A/H3N2 and influenza B.

*Inference method*

Only a fraction of individuals presenting with ILI/ARI symptoms are infected with influenza. We multiplied the weekly rates of ILI/ARI consultation per 10,000 individuals [15,17] by the weekly percentages of respiratory specimens from symptomatic individuals that were RT-PCR positive for each of influenza A/H1N1, A/H3N2 and B [15,17] to estimate the weekly incidence proxies for each of the corresponding influenza (sub)types:

*Weekly influenza (sub)type incidence proxy =* (1)

*= Rate of consultations for ILI/ARI * % All respiratory specimens that were positive for the (sub)type*

As noted in [2], those proxies are expected to be *proportional* to the weekly population incidence for the each of the major influenza (sub)types (hence the name "proxy"). Monthly incidence proxies for influenza A/H1N1, A/H3N2 and B were obtained as the weighted average of the weekly incidence proxies for those weeks that overlapped with a given month; specifically, for each influenza (sub)type and month, the incidence proxy for each week was multiplied by the number of days in that week that were part of the corresponding month (e.g. 7 if the week was entirely within that month), then the results were summed over the different weeks and divided by the number of days in the corresponding month. Figure 1 plots the monthly incidence proxies (which are estimated as daily averages over a month) for influenza A/H1N1, A/H3N2 and B in Russia between 09/2010 and 12/2015 (64 months), with that time period chosen for inference purposes as we have information on both influenza surveillance and monthly mortality during that period [15-17]. Figure 1 suggests a very large incidence proxy for influenza A/H1N1 during the 2010/11 season. This may have to do with the fact that the age distribution of cases for the 2010/11 A/H1N1 epidemic was different compared to the subsequent A/H1N1 epidemics, and the ratio between the A/H1N1 incidence proxy and the rates of A/H1N1-associated mortality is potentially different for the 2010/11 season compared to later seasons. Correspondingly, we split the A/H1N1 incidence proxy into two ($A/H1N1^1$ for the 2010/11 season, and $A/H1N1^2$ for the later seasons).

Earlier work (e.g. [2]) suggests that there is a delay between one and two weeks between influenza incidence and associated mortality. For the purposes of relating influenza incidence to mortality we considered influenza incidence proxies shifted forward by either 1 or 2 weeks, with the shifted monthly incidence proxies derived from the shifted weekly incidence proxies in a manner described in the preceding paragraph. For example, the incidence proxy for a given month shifted by 2 weeks is the weighted average of the weekly incidence proxies for those weeks whose 2-week *forward* shift overlaps with a given month. The choice of shifting the influenza incidence proxies by 1 week or 2 weeks in fitting the mortality data is determined by the Akaike Information Criterion (AIC) scores similarly to other choices of parameters described below.

If $M(t)$ is the mortality rate for a given cause on month $t$ (with $t = 1$ for 09/2010), and $A/H3N2(t), A/H1N1^1(t), A/H1N1^2(t), B(t)$ are the shifted incidence proxies for the major influenza (sub)types on month $t$, then the inference model in [2,3] suggests that

$$M(t) = \beta_0 + \beta_1 \cdot A/H3N2(t) + \beta_2 \cdot A/H1N1^1(t) + \beta_3 \cdot A/H1N1^2(t) + \beta_4 \cdot B(t) +$$
$$Baseline + Trend + + Noise \qquad (2)$$

Here *Baseline* is the baseline rate of mortality not associated with influenza circulation that is *periodic* with yearly periodicity. To accommodate the baseline's unknown shape, we will model it as

$$Baseline(t) = \beta_5 \cdot \cos\left(\frac{2\pi t}{12}\right) + \beta_6 \cdot \sin\left(\frac{2\pi t}{12}\right) + \beta_7 \cdot \cos\left(\frac{2\pi t}{6}\right) + \beta_8 \cdot \sin\left(\frac{2\pi t}{6}\right) + \beta_9 \cdot Jan$$

Here *Jan* is a variable equaling 1 for the month of January, 0 otherwise (see Supporting Information). The reason for including this variable is that data suggest bumps in mortality in January, even under low influenza circulation/ ILI/ARI rates (particularly for the 2012/13 season). Such bumps (also seen for the 2007/08 season under low influenza circulation) are likely explained by cold weather [6] (with the January effect on circulatory mortality in Russia examined in [18]), and possibly by the circulation of other respiratory viruses. The (temporal) trend is modeled as a low degree polynomial in time, or time periods, as explained below, with explicit model equations for respiratory mortality, circulatory mortality and mortality for certain infectious and parasitic diseases presented further down in this section.

The Akaike Information Criterion (AIC) is adopted to select the variables (covariates) in the model for each mortality cause, with the variable whose omission results in the largest decline in the AIC score being dropped at each step until no omissions of a variable decrease the AIC score. The model equations for respiratory mortality, circulatory mortality and mortality due to certain infectious and parasitic diseases are as follows:

*Respiratory mortality.* While there is no obvious long-term trend in respiratory mortality in Figure 2, clearly there are seasonal fluctuations not related to influenza circulation, and those are likely related to the circulation of other respiratory viruses. We will accommodate those seasonal effects by modeling the temporal trend as a function of a season, where the season for the circulation of other respiratory viruses is modeled to run from July to June. Let $Ts1$ be a linear function of the season (equaling 1 for the months of 09/2010 through 06/2011, to 2 for 07/2011 through 06/2012 etc.), $Ts2 = Ts1^2, Ts3 = Ts1^3$, etc. are the powers in $Ts1$. Then the model equation for respiratory mortality selected using the AIC criterion is:

$$M_{resp}(t) = \beta_0 + \beta_1 \cdot A/H3N2(t) + \beta_2 \cdot A/H1N1^1(t) + \beta_3 \cdot A/H1N1^2(t) + \beta_4 \cdot B(t) + \beta_5 \cdot \sin\left(\frac{2\pi t}{12}\right) + \beta_6 \cdot \cos\left(\frac{2\pi t}{6}\right) + \beta_7 \cdot Ts2 + \beta_8 \cdot Ts3 + \beta_9 \cdot Ts4 + \beta_{10} \cdot Jan \quad (3)$$

We note that all the variables in the linear regression equation (3) are essential in the sense that removing each of them increases the AIC score of the model.

*Circulatory mortality and mortality due to certain infectious and parasitic diseases.* Figure 2 suggests that there is a continuous downward trend in those two mortality time series. Therefor, we model the trend as a low degree polynomial in the month $t$, not the season (with that choice justified not only by the visual inspection of Figure 2 but also by the AIC criterion). The model equations selected by the AIC criterion are:

Circulatory deaths: $M_{circ}(t) = \beta_0 + \beta_1 \cdot B(t) + \beta_2 \cdot \cos\left(\frac{2\pi t}{12}\right) + \beta_3 \cdot t^2 + \beta_4 \cdot t^3 + \beta_5 \cdot Jan$ (4)

Infectious and Parasitic disease deaths: $M_{IP}(t) = \beta_0 + \beta_1 \cdot A/H1N1^1(t) + \beta_2 \cdot B(t) + \beta_3 \cdot \cos\left(\frac{2\pi t}{12}\right) + \beta_4 \cdot \sin\left(\frac{2\pi t}{12}\right) + \beta_5 \cdot t^2 + \beta_6 \cdot t^3 + \beta_7 \cdot t^4$ \quad (5)

We note that a number of terms for the different influenza (sub)types are omitted by the AIC criterion for the mortality causes in eqns. (4,5). Compared to respiratory mortality, those omissions might be related to the fact that the relative effect of influenza circulation on mortality is higher for respiratory deaths compared to the other categories of deaths; correspondingly, the effects of circulation of each influenza

(sub)type on the monthly mortality are mixed with the relatively larger fluctuations in the monthly mortality counts for non-respiratory categories compared to the respiratory categories, resulting in potential lack of ascertainability as expressed by the (lack of) justification of including those covariates via the AIC score. We also note in that regard that it is visually clear that for example the incidence proxy for influenza A/H3N2 was growing towards its peak when circulatory mortality was declining for three out of five seasons in the data, which may help explain its exclusion from the corresponding model. Given the high correlation between the incidence proxies for the different influenza (sub)types, omitting one (sub)type from the regression model affects the regression coefficients for the other (sub)types. Thus models given by eqns. (4,5) are meant to estimate the contribution of all influenza (sub)types, not only the ones included to the corresponding mortality rates, though only as the average over the study period (with annual estimates being less certain) – see also the Discussion regarding the contribution of influenza A to circulatory mortality in Russia.

## Results

Figure 1 plots the monthly incidence proxies for influenza A/H1N1, A/H3N2 and B in Russia between 09/2010 and 12/2015. Figure 2 plots the monthly rates of respiratory deaths, circulatory deaths, and deaths for certain infectious and parasitic diseases between 09/2010 and 12/2015. One sees that during seasons of high influenza circulation (particularly the 2014/15 season that had the highest circulation of influenza B and A/H3N2, Figure 1), the peak periods of each of the three mortality times series in Figure 2 were largest; for the relatively smaller influenza seasons (2011/12 and 2013/14), peaks of the corresponding mortality time series were relatively smaller.

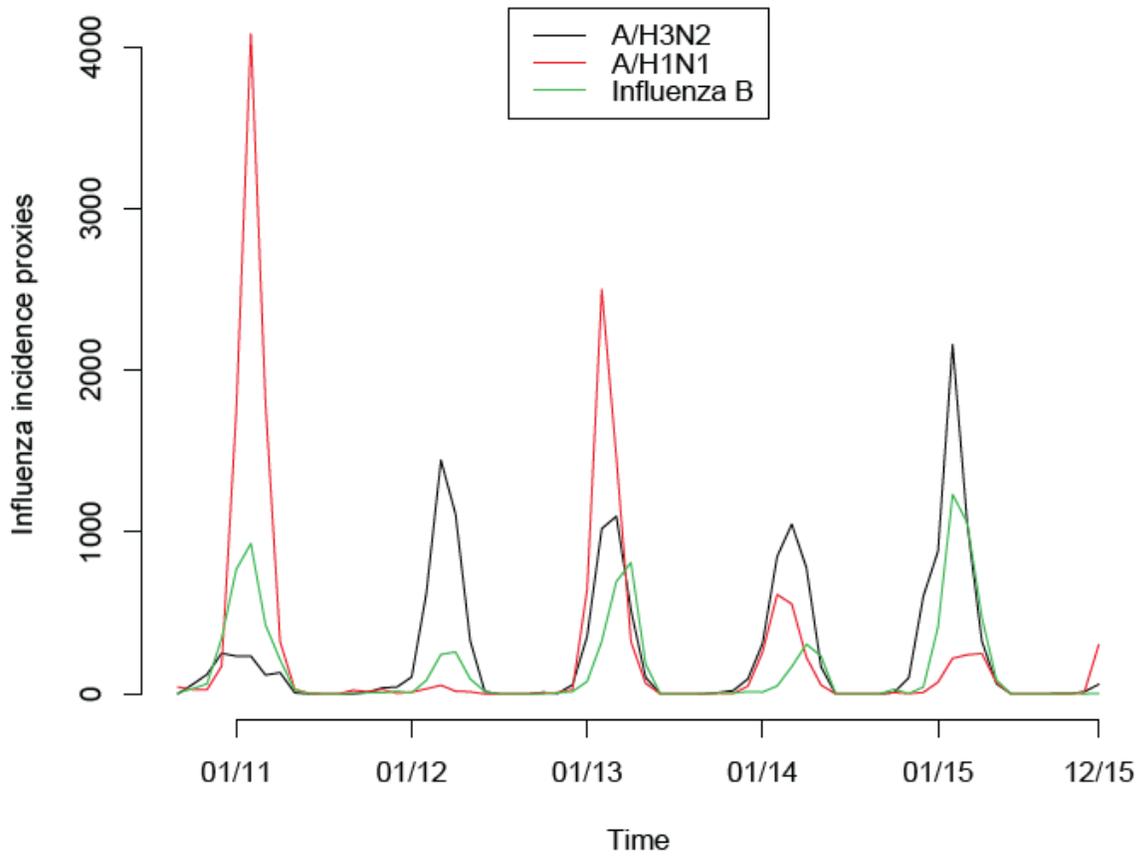

**Figure1:** Monthly incidence proxies for influenza A/H3N2, A/H1N1 and B in Russia, 9/2010 through 12/2015.

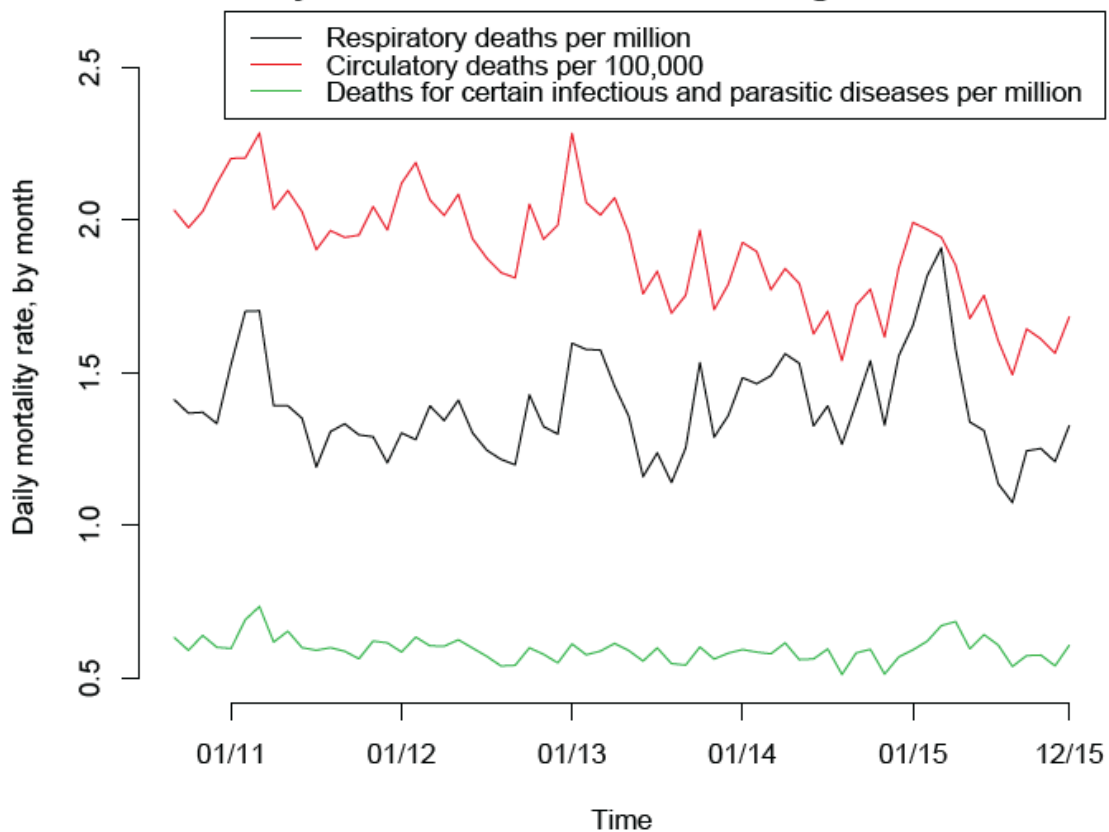

**Figure 2:** Rates of daily mortality for respiratory causes (per one million individuals in Russia), circulatory causes (per 100,000 individuals), and for certain infectious and parasitic diseases (per one million individuals) by month, 9/2010 through 12/2015.

*Respiratory deaths.* Figure 3 plots the fits for the model for respiratory mortality given by eq. (3). The fits appear to be fairly temporally consistent, with the weekly contribution of influenza to mortality estimated as the difference between the red and the green curves in Figure 3.

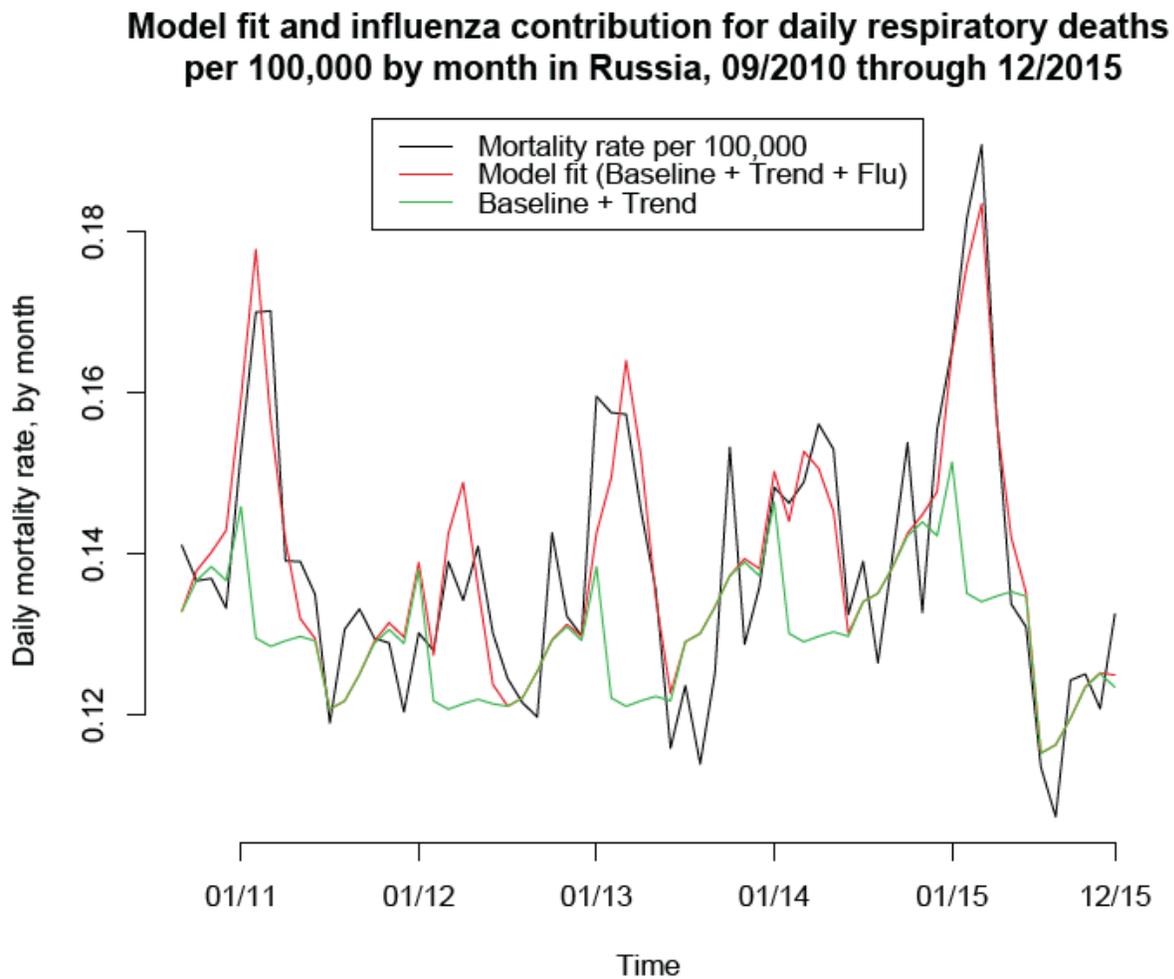

**Figure 3:** Model fit for the daily rates of respiratory deaths per 100,000 by month in Russia, 9/2010 though 12/2105. The contribution of influenza to respiratory mortality is represented by the difference between the red and the green curves.

Table 1 presents the estimates of the contribution of influenza A/H3N2, A/H1N1, influenza B and all of influenza to respiratory mortality for each of the 2010/11 through the 2014/15 seasons (with each influenza season assumed to run from September through June), as well as the annual averages for the corresponding contributions for the 2010/11 through the 2014/15 time period (5 seasons), with the nascent 2015/2016 influenza epidemic between 09/2015 and 12/2015 excluded from the latter estimates (though data for the 09/2015 through the 12/15 period is included in the regression model). We estimate an annual average of 4552(3744,5360) influenza-associated respiratory deaths during the study period, with over a half of those deaths being due to influenza A/H3N2. The 2014/15 season had the largest number of estimated influenza-associated deaths: 6128(4734,7521).

| Season /Flu type | A/H3N2 | A/H1N1 | Influenza B | All influenza |
|---|---|---|---|---|
| 2010-11 | 837(438,1237) | 2304(570,4039) | 1687(73,3300) | 4828(3419,6237) |
| 2010-12 | 2722(1423,4021) | 61(5.1,116) | 445(19,872) | 3228(2198,4258) |
| 2012-13 | 2286(1195,3377) | 1554(131,2976) | 1287(56,2517) | 5126(3781,6471) |
| 2013-14 | 2402(1255,3549) | 554(47,1062) | 495(21,969) | 3451(2621,4281) |
| 2014-15 | 3807(1990,5624) | 279(24,535) | 2042(88,3995) | 6128(4734,7521) |
| Annual average | 2411(1260,3561) | 950 (374,1527) | 1191(52,2331) | 4552(3744,5360) |

**Table 1:** Seasonal and average annual number of respiratory deaths associated with the different influenza (sub)types, as well as with influenza overall for the 2010-11 through the 2014-15 seasons in Russia.

*Circulatory deaths and deaths due to certain infectious and parasitic diseases.* Table 2 presents the estimates of the average annual numbers of circulatory deaths, as well as deaths due to certain infectious and parasitic diseases associated with influenza circulation for the 2010/11 through the 2014/15 seasons. We estimate an annual average of 11106 (4321,17892) influenza-associated circulatory deaths, and an annual average of 343 (63,624) influenza-associated deaths due to certain infectious and parasitic diseases. We note that the estimated ratio of circulatory to respiratory influenza-associated deaths in Russia (about 2.44-to-1 Table 2 vs. Table 1) is quite higher that the corresponding estimate (about 1.35-to-1, [2]) for the US. At the same time, the overall ratio of circulatory-to-respiratory deaths in Russia (about 14-to-1 in 2018, [12]) is much higher than the corresponding ratio in the US (about 3.08-to-1 in 2017, [19]).

| Cause | Annual average number of deaths |
|---|---|
| Circulatory deaths | 11106 (4321,17892) |
| Deaths for certain infectious and parasitic diseases | 343 (63,624) |

**Table 2:** Average annual number of influenza-associated deaths for circulatory causes and for certain infectious and parasitic diseases for the 2010-11 through the 2014-15 seasons in Russia.

Figures 4,5 plot the model fits for the rates of circulatory deaths, as well as deaths due to certain infectious and parasitic diseases. The model fit for circulatory deaths is less consistent than for respiratory deaths (Figure 4 vs. 3). For example, the major circulatory mortality peak during the 2012/13 season is not well-explained by the model in eq. (4). At the same time, this peak might indeed largely be not influenza-associated as it preceded the period of high influenza circulation as described by the incidence proxies in Figure 1, and this also applies to an extent for the 2011/12 season – see also the 2nd paragraph of the Discussion.

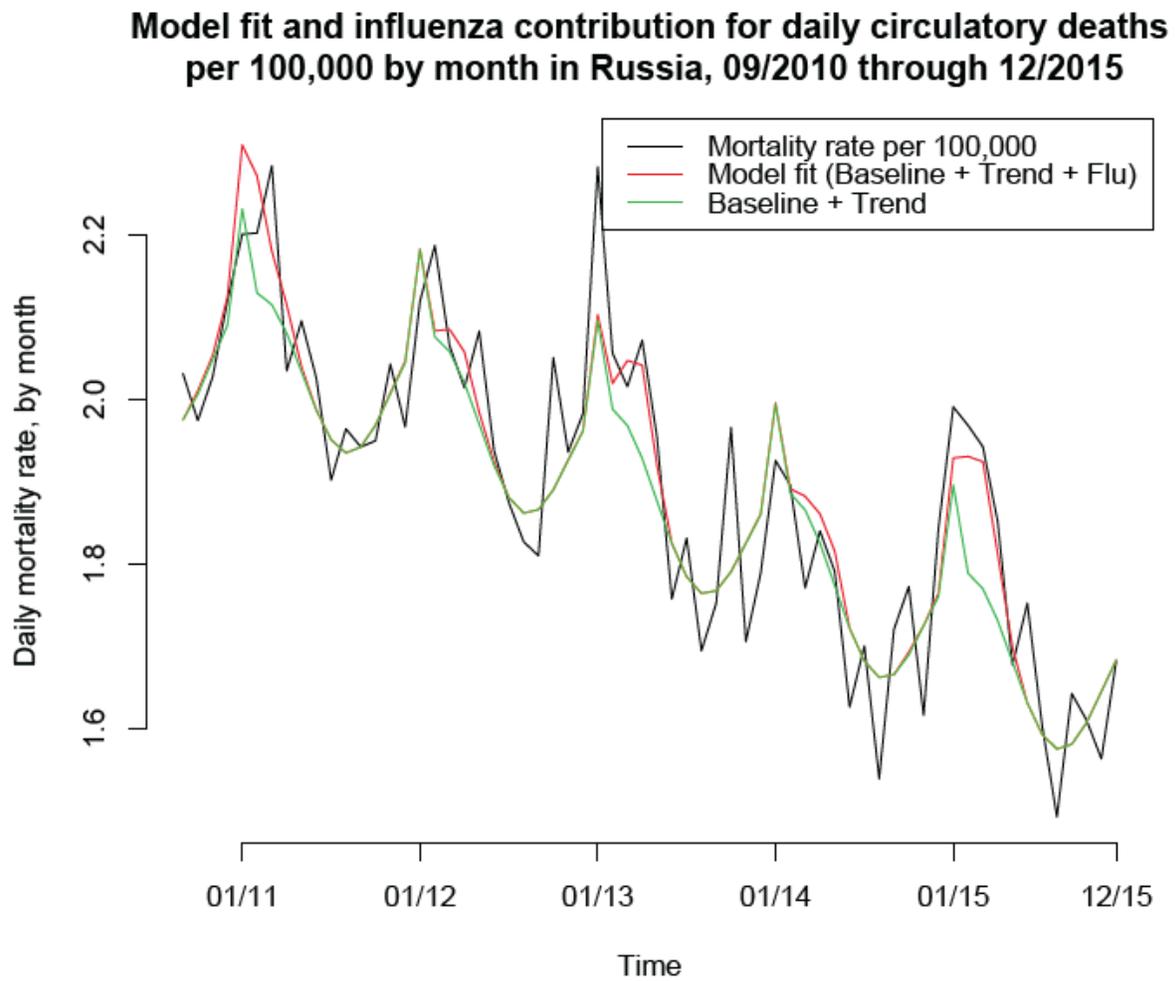

**Figure 4:** Model fit for the daily rates of circulatory deaths per 100,000 by month in Russia, 9/2010 though 12/2105. The contribution of influenza to circulatory mortality is represented by the difference between the red and the green curves.

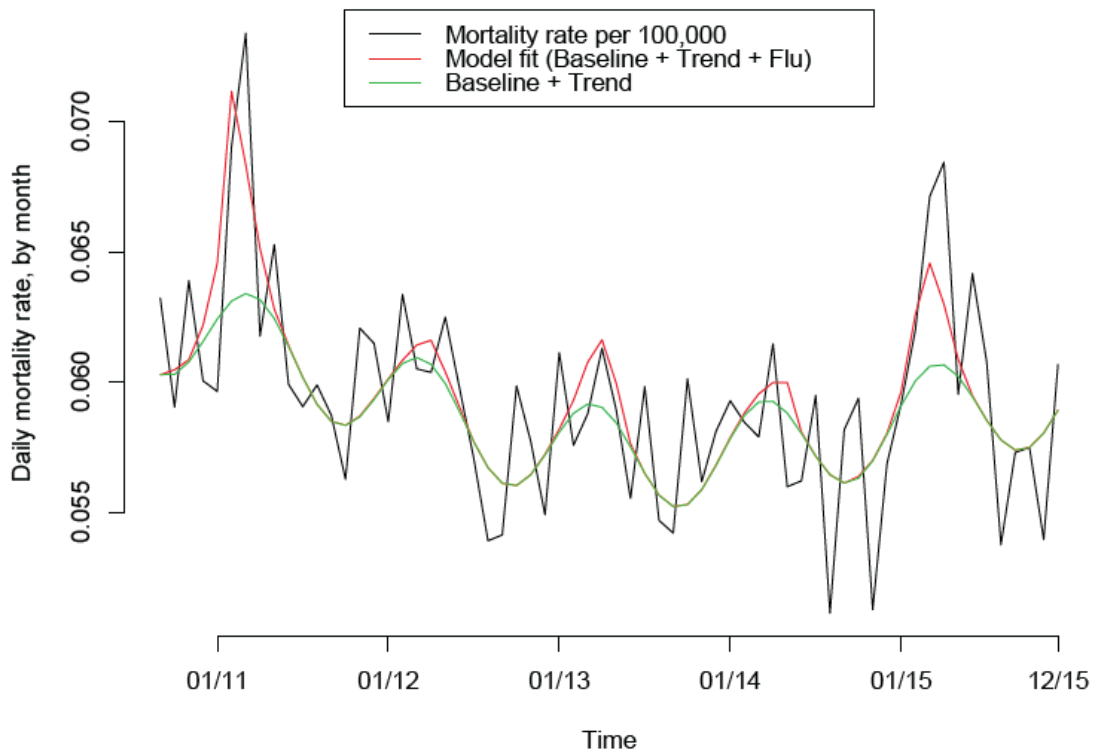

**Figure 5:** Model fit for the daily rates of deaths for certain infectious and parasitic diseases per 100,000 by month in Russia, 9/2010 though 12/2105. The contribution of influenza to mortality for certain infectious and parasitic diseases is represented by the difference between the red and the green curves.

**Discussion**

Influenza circulation is known to be associated with substantial mortality burden in the Northern Hemisphere, e.g. [2-7]. At the same time, there is limited information on the volume of influenza-related mortality in Russia, though some regional estimates, particularly for the city of St. Petersburg are available [9]. In this paper, we applied the previously developed methodology that was used to evaluate the rates of influenza-associated mortality in several countries, e.g. [2,3,5-7] to estimate the rates of influenza-associated mortality for three causes of death (respiratory causes, circulatory causes, and certain infectious and parasitic diseases) for the period between the 2010/11 through the 2014/15 influenza seasons in Russia. We estimate about 16,001 annual influenza-associated deaths for those three causes during that period, the majority of which (about 11,106) are for circulatory causes, with influenza A/H3N2 being responsible for the majority of influenza-associated respiratory deaths and influenza B playing a substantial role in the influenza-associated circulatory mortality (see also [13]). The prominence of circulatory deaths among influenza-associated mortality may be related to the overall high rates of circulatory deaths in Russia [12]. We hope that this work may help inform mitigation efforts. In particular, the role played by influenza B in the circulatory mortality burden supports the potential benefit of quadrivalent influenza vaccines (with trivalent influenza vaccines not containing the influenza B/Yamagata strains that can be a source of high levels of excess mortality [10]). We also note in that regard that the 2018/19 influenza season saw minimal circulation of influenza B in Russia [15], raising the possibility of a major influenza B epidemic during the 2019/20 season, which further supports the need for a quadrivalent influenza vaccine, particularly for individuals with cardiac disease. Additionally, administration of antiviral medications for certain population subgroups defined by age and presence of underlying health conditions (particularly circulatory diseases) may help mitigate the burden of severe outcomes during influenza epidemics in Russia.

Our study has some limitations. We only had access to monthly mortality data, with the signal of influenza circulation in those data being mixed with other fluctuations of relatively high magnitude, particularly for non-respiratory causes. While the model fits for respiratory causes exhibited a good deal of temporal consistency, this wasn't the case for circulatory causes during certain seasons, particularly the 2012/13 season. While earlier work, e.g. [13], as well as Table 3 in [2] (comparison of regression coefficients for circulatory vs. respiratory deaths) also supports the important role played by influenza B in circulatory influenza-associated mortality, there is uncertainty regarding our estimates of the contribution of influenza A to circulatory mortality. It might be that our model had missed some of the contribution of influenza, particularly influenza A to mortality, especially circulatory mortality, with the circulatory mortality peaks during the 2012/13 season potentially being related to influenza A infections contrary to the model's results (Figure 4). At the same time, other factors might have affected this peak in circulatory mortality. Indeed, for the 2012/13 season, the peak of circulatory mortality preceded the peak periods of both influenza circulation and ILI/ARI rates. Moreover, the peak of respiratory mortality was broader than the peak of circulatory mortality during that season (Figure 2), with the later part of that peak in respiratory mortality explained well by influenza circulation (Figure 3), and the earlier part, matching temporally the peak in circulatory mortality (Figure 2), not being explained by influenza circulation (Figure 3). It may be that cold weather [6], and possibly the circulation of other respiratory viruses have affected both the respiratory and circulatory mortality during the 2012/13 season. We also note the major peak in both circulatory and respiratory mortality in January, 2008 that couldn't be explained by influenza circulation (Figures 1 and 4 in [1]). Further work involving more granular data, including climatic data [18], weekly mortality data, as well as data stratified by age is needed to better understand "excess mortality" and the contribution of influenza to the mortality rates in Russia, particularly the contribution of influenza A to circulatory mortality.

We believe that despite those limitations, this work provides a national study of influenza-related mortality in Russia, giving evidence for the substantial burden of influenza-associated mortality, particularly for circulatory causes, and the contribution of influenza B to the volume of influenza-associated mortality. We hope that this work would stimulate further efforts involving more granular data (in particular, weekly mortality data, as well as data stratified by age) to better understand the effect of influenza epidemics in Russia on mortality and help inform mitigation efforts for certain population subgroups.

**References**

[1] Sominina A, Burtseva E, Eropkin M, Karpova L, Zarubaev V, Smorodintseva E, et al. INFLUENZA SURVEILLANCE IN RUSSIA BASED ON EPIDEMIOLOGICAL AND LABORATORY DATA FOR THE PERIOD FROM 2005 TO 2012. Am J Infect Dis. 2013;9(3):77-93

[2] Goldstein E, Viboud C, Charu V, Lipsitch M. Improving the estimation of influenza-related mortality over a seasonal baseline. Epidemiology. 2012;23(6):829-38.

[3] Quandelacy TM, Viboud C, Charu V, Lipsitch M, Goldstein E.  Age- and Sex-related Risk Factors for Influenza-associated Mortality in the United States Between 1997-2007. Am J Epidemiol. 2014;179(2):156-67

[4] Thompson WW, Shay DK, Weintraub E, et al. Mortality associated with influenza and respiratory syncytial virus in the United States. JAMA. 2003;289(2):179–186

[5] Wu P, Goldstein E, Ho LM, Yang L, Nishiura H, Wu JT, et al.  Excess mortality associated with influenza A and B virus in Hong Kong, 1998-2009. J Infect Dis. 2012;206(12):1862-71

[6] Pebody RG, Green HK, Warburton F, Sinnathamby M, Ellis J, Mølbak K, et al. Significant spike in excess mortality in England in winter 2014/15 - influenza the likely culprit. Epidemiol Infect. 2018;146(9):1106-1113

[7] Nielsen J, Krause TG, Mølbak K. Influenza-associated mortality determined from all-cause mortality, Denmark 2010/11-2016/17: The FluMOMO model. Influenza Other Respir Viruses. 2018;12(5): 591–604.

[8] Mamas MA, Fraser D, Neyses L. Cardiovascular manifestations associated with influenza virus infection. Int J Cardiol. 2008;130(3):304-9.

[9] L.S. Karpova, K.M. Volik, K.A. Stolyarov, N.M. Popovtseva, T.P. Stolyarova. Excess mortality from separate nosological forms of somatic and infectious diseases among children and adults in Saint Petersburg since 2009 to 2015. Population Health 2016;5:39-44

[10] Vestergaard LS, Nielsen J, Krause TG, Espenhain L, Tersago K, Bustos Sierra N, et al. Excess all-cause and influenza-attributable mortality in Europe, December 2016 to February 2017. Euro Surveill. 2017;22(14). pii: 30506

[11] Nielsen J, Vestergaard LS, Richter L, Schmid D, Bustos N, Asikainen T, et al. European all-cause excess and influenza-attributable mortality in the 2017/18 season: should the burden of influenza B be reconsidered? Clin Microbiol Infect. 2019;pii: S1198-743X(19)30058-8.

[12] Russian Federal State Statistics Service (Rosstat). Number of deaths according to different causes (in Russian). 2019. Available from the following database: http://www.gks.ru/wps/wcm/connect/rosstat_main/rosstat/ru/statistics/population/demography/#

[13] Kwong JC, Schwartz KL, Campitelli MA, Chung H, Crowcroft NS, Karnauchow T, et al. Acute Myocardial Infarction after Laboratory-Confirmed Influenza Infection. N Engl J Med. 2018;378(4):345-353

[14] Goldstein E, Finelli L, O'Halloran A, Liu P, Karaca Z, Steiner CA, et al. Hospitalizations associated with respiratory syncytial virus (RSV) and influenza in children, including children diagnosed with asthma. Epidemiology 2019 Aug 12. doi: 10.1097/EDE.0000000000001092. [Epub ahead of print]

[15] Smorodintsev Research Institute of Influenza (RII). Influenza Surveillance System in Russia / Epidemic Situation. 2019. Available from:

https://www.influenza.spb.ru/en/influenza_surveillance_system_in_russia/epidemic_situation/

[16] Russian Federal State Statistics Service (Rosstat). Monthly number of deaths for certain major causes (in Russian). 2019. Available from the following database: http://cbsd.gks.ru/

[17] Smorodintsev Research Institute of Influenza (RII). Influenza Surveillance System in Russia / Epidemic Situation (in Russian). 2019. Available from: https://www.influenza.spb.ru/system/epidemic_situation/situation_on_a_flu/

[18] Hansulin VI, Gafarov VV, Voevoda VI, Artamonova MV. Indicators of mortality for circulatory diseases in relation to annual temperature and latitude in the Russian Federation. International Journal of Applied and Fundamental Studies. 2015;6(2):255-259 (in Russian) https://applied-research.ru/ru/article/view?id=6883

[19] US CDC Wonder. 1999-2017 Multiple Cause of Death Request. 2019. Available from: https://wonder.cdc.gov/mcd-icd10.html

Supporting Information for: **Influenza-associated mortality for different causes of death during the 2010-2011 through the 2014-2015 influenza seasons in Russia**

Here, we examine the results of the inference model when the variable for the January effect of mortality is not included. Below are the equations for the models (selected by the AIC criterion) for respiratory mortality, circulatory mortality, and mortality for certain infectious and parasitic diseases (compare with eqns. 3-5):

Respiratory: $M_{resp}(t) = \beta_0 + \beta_1 \cdot A/H3N2(t) + \beta_2 \cdot A/H1N1^1(t) + \beta_3 \cdot A/H1N1^2(t) + \beta_4 \cdot B(t) + \beta_5 \cdot \cos\left(\frac{2\pi t}{12}\right) + \beta_6 \cdot \sin\left(\frac{2\pi t}{12}\right) + \beta_7 \cdot Ts2 + \beta_8 \cdot Ts3 + \beta_9 \cdot Ts4$    (S1)

Circulatory: $M_{circ}(t) = \beta_0 + \beta_1 \cdot B(t) + \beta_2 \cdot \cos\left(\frac{2\pi t}{12}\right) + \beta_3 \cdot t^2 + \beta_4 \cdot t^3$    (S2)

Infectious and Parasitic diseases: $M_{IP}(t) = \beta_0 + \beta_1 \cdot A/H1N1^1(t) + \beta_2 \cdot B(t) + \beta_3 \cdot \cos\left(\frac{2\pi t}{12}\right) + \beta_4 \cdot \sin\left(\frac{2\pi t}{12}\right) + \beta_5 \cdot t^2 + \beta_6 \cdot t^3 + \beta_7 \cdot t^4$    (S3)

Table S1 summarizes the comparison for the models with and without the January effect in terms of the AIC score, R-squared statistic, and estimated average annual numbers of influenza-associated deaths. One sees significant improvement in the model fits for respiratory and circulatory deaths with the inclusion of the January effect – see also [S1].

|  | Model with January effect | | | Model without January effect | | |
| --- | --- | --- | --- | --- | --- | --- |
|  | AIC score | R-squared | Average annual mortality | AIC score | R-squared | Average annual mortality |
| Respiratory | -422.28 | 0.795 | 4552 (3744,5360) | -416.76 | 0.769 | 3659 (2475,4843) |
| Circulatory | -143.06 | 0.854 | 11106 (4321,17892) | -136.76 | 0.833 | 9558 (2280,16836) |
| Certain infectious parasitic diseases | -561.26 | 0.563 | 343 (63,624) | -561.26 | 0.563 | 343 (63,624) |

**Table S1:** Comparison of the models with and without the January effect on mortality.

**References**

[S1] Hansulin VI, Gafarov VV, Voevoda VI, Artamonova MV. Indicators of mortality for circulatory diseases in relation to annual temperature and latitude in the Russian Federation. International Journal of Applied and Fundamental Studies. 2015;6(2):255-259 (in Russian) https://applied-research.ru/ru/article/view?id=6883